\documentclass[9pt,twocolumn]{article}
\usepackage{epsfig}
\begin{document}
\newcommand{\vcii}[2]{
   \left(
      \begin{array}{c}
         #1\\
         #2
      \end{array}
   \right)
}
\newcommand{\D}{\displaystyle}
\newcommand{\T}{\textstyle}
\newcommand{\SC}{\scriptstyle}
\newcommand{\SY}{\scriptscriptstyle}

\begin{minipage}{18cm}
\begin{center}

\vspace{0.5cm}

{\large \bf A Bicharacteristic Scheme for the Numerical Computation of\\
Two-Dimensional Converging Shock Waves}\\
\vspace{0.5cm}
 
U. E. Meier and F. Demmig\\
Institut f\"ur Atom- und Molek\"ulphysik, Abteilung Plasmaphysik\\
Universit\"at Hannover, Germany\\
Phone: +49-511-762-3543, Fax:   +49-511-762-2784\\
email: demmig@pmp.uni-hannover.de\\
\end{center}
\end{minipage}\\

\vspace{0.7cm}

\noindent
{\bf ABSTRACT\\

A 2d unsteady bicharacteristic scheme with shock fitting is presented and
its characteristic step, shock point step and boundary step are described.
The bicharacteristic scheme is compared with an UNO scheme 
and the Moretti scheme. Its capabilities are illustrated by computing a 
converging, deformed shock wave.}\\

\noindent
{\bf INTRODUCTION\\}

Commonly used difference schemes are capable of computing complex
flow patterns and shock configurations,
but they smear out shocks and other discontinuities. To avoid the
effects of shock smearing, shock fitting has to be used. 
This is most naturally done by the method of bicharacteristics.\\

\noindent
{\bf \underline{Characteristic Normal Form}}

Introducing the substantial derivate 
$D_0=D_t=\partial_t +\left(\vec v \nabla \right)$,
a directional derivative operator 
$D_B:= (\vec B\nabla)= \partial_t + (( \vec v-a\vec g) \nabla)$ along a bicharacteristic,
the abbreviations $v_g:=\vec v\vec g$, $\partial_g:=\vec g\nabla$
and the logarithmic pressure $P$ and density $\varrho$,
the Euler equations can be written as\\

\noindent
\[\begin{array}{rll}
   \D D_t \varrho &= \frac{1}{\gamma}D_t P       &\mbox{\rm energy equation}\\
   \D D_t P       &= -\gamma \nabla \vec v       &\mbox{\rm continuity equation}\\
   \D D_t \vec v  &= -\frac{a^2}{\gamma}\nabla P &\mbox{\rm momentum equation}\\
  \end{array}
\]

or, in characteristic normal form:

\vspace{3mm}

\noindent
\[\begin{array}{|rl|c|}\hline
   \multicolumn{2}{|c|}{\mbox{\rule[-3mm]{0mm}{8mm}\rm characteristic differential equation}} & \multicolumn{1}{c|}{\mbox{\rule[-3mm]{0mm}{8mm}\rm direction}}\\
   \D D_0 \varrho &\D = \frac{1}{\gamma}D_0 P &\vcii{1}{\vec v}\\[6mm]
   \D D_B v_g  -\frac{a}{\gamma}D_B P &\D = a (\nabla \vec v-\partial_gv_g) &\vcii{1}{\vec v-a\vec g}\\[6mm]\hline
  \end{array}
\]

\vspace{3mm}

\noindent
where the characteristic differential equations (CDEs) are valid 
only on their specific characteristic surfaces, i.e. the particle path 
$\vec C_0=(1,\vec v)$ and the Monge cone, generated by the 
bicharacteristics $\vec B=(1,\vec v -a\vec g)$. The unit vector $\vec g$ 
singles out a specific bicharacteristic on the Monge cone 
(fig. 1).\\

\begin{figure}[t]
\unitlength1cm
\begin{picture}(8,4)
\end{picture}
\end{figure} 

\vspace{1cm}

\noindent
{\bf \underline{Characteristic Step}}

The bicharacteristic scheme was constructed on a Cartesian grid, 
in analogy to Hartree's method [1].
Using an educated guess of $\vec v$ and $a$ in a grid point at the 
new time level the bicharacteristics are drawn backwards 
in the directions of the coordinate lines.  
At the intersection points of these bicharacteristics with the old time 
level, called 'footpoints', the quantities P, $\rho$, $\vec v$ are 
interpolated in second order from the grid points with a 2d
nine point least-squares method (fig. 1). 
Using the discretized CDEs, the quantities P, $\rho$, $\vec v$ 
then can be determined in first order at the new time level.
A corrector step, which uses the first order values of $\vec v$ 
and $a$ to reconstruct the bicharacteristics, yields the quantities P, 
$\rho$, $\vec v$ in second order, eliminating the lateral derivatives 
on the new time level according to Butler [2].\\

\begin{figure}[h] 
\epsfclipon
\centerline{\epsfxsize 6cm \epsfbox{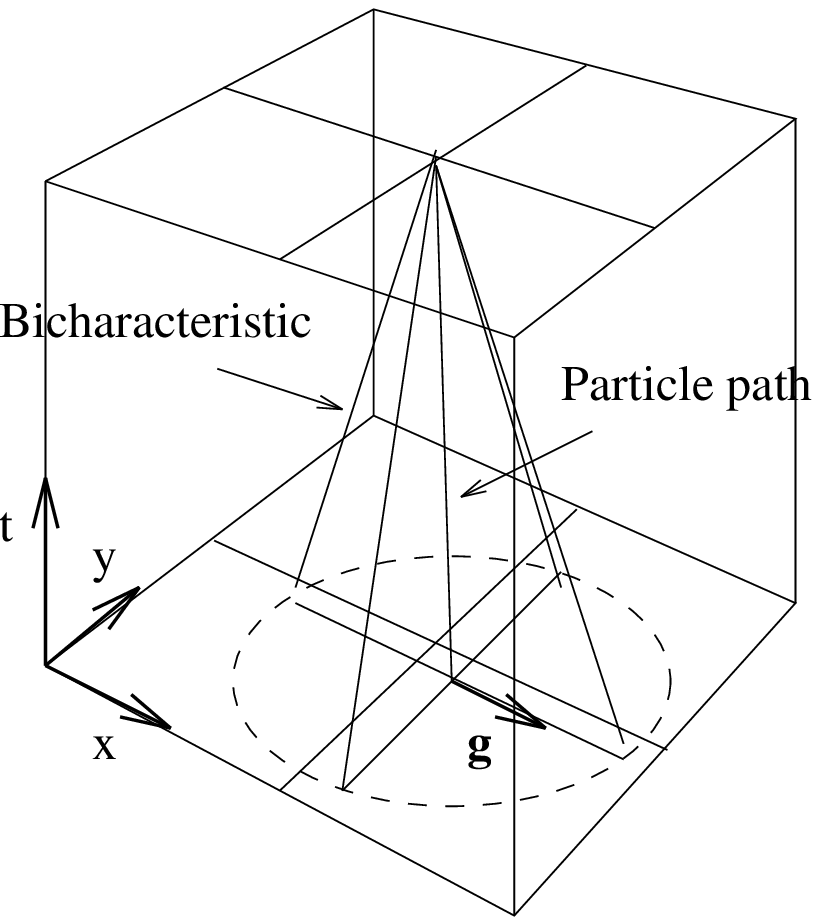}}
\centerline{Figure 1: Characteristic Step}
\end{figure}

\noindent
{\bf \underline{Treatment Of Shocks}}

To begin with, a shock point is moved using its previous velocity. 
At its new location, the (oblique) Rankine-Hugoniot conditions 
are used to determine the downstream values. 
With these values, a bicharacteristic is drawn backwards antiparallel
to the shock front normal (fig. 2). 
The correct position of the new shock point is found by iteration, 
such that both the Rankine-Hugoniot conditions and the 
differential equation along the bicharacteristic are fullfilled.

  Shock points are moved along their trajectories. 
  The density of shock points on a shock front can be kept constant  
  by creating or deleting shock points. Two pointers are attached 
  to each shock point, pointing to its predecessor and successor. 
  Though all shock points are stored in an arbitrary sequence in a single 
  array, shock contours and their normal vectors can nevertheless be 
  reconstructed with the help of the  
  pointers. New shock points are detected with an algorithm according to 
  Moretti [3] at each new time level.\\

\begin{figure}[h] \epsfclipon
\centerline{\epsfxsize 6cm \epsfbox{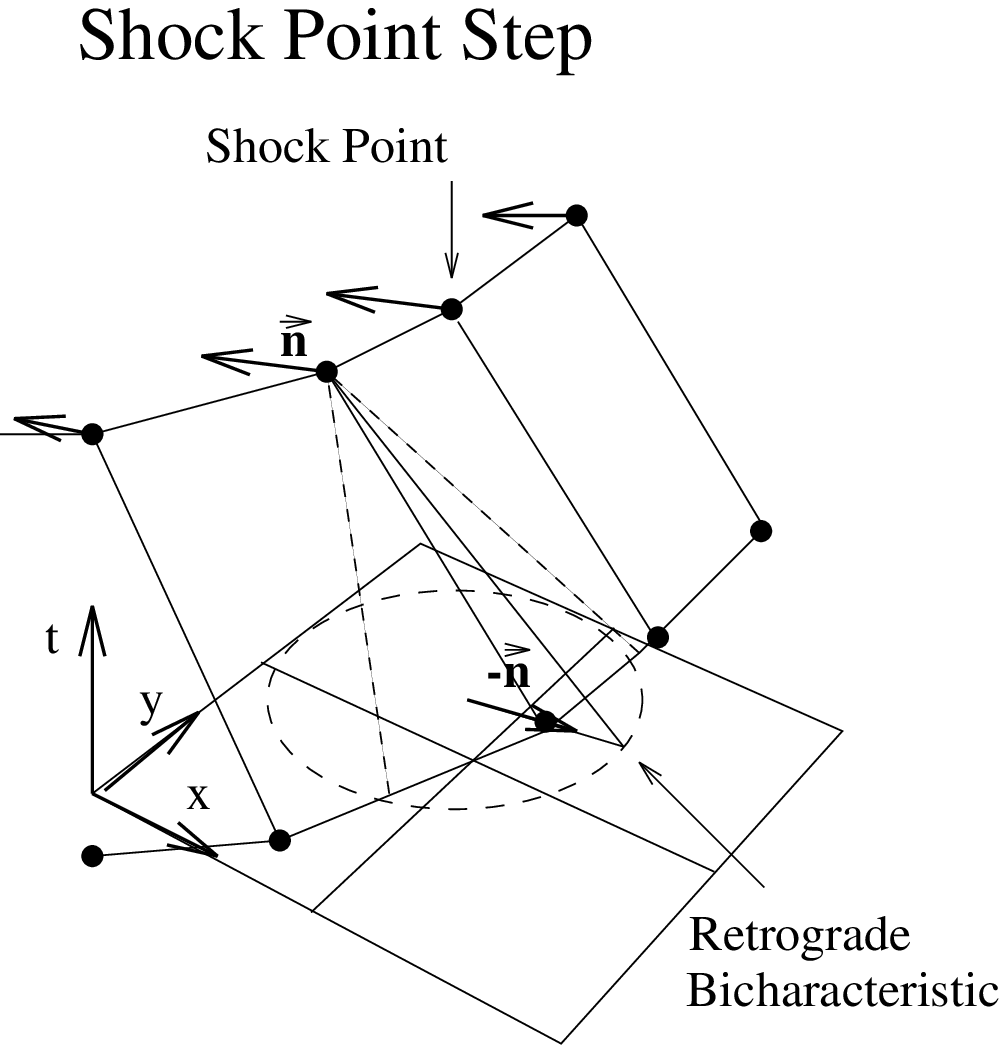}}
\centerline{Figure 2: Shock Point Step}
\end{figure}

\noindent
{\bf \underline{Boundary Step}}

When a new time level is computed, it is first done regardless of all 
discontinuities. 
Then, the shock points are moved. 
Invalid grid points, which were computed by drawing bicharacteristics crossing
a shock surface, are updated afterwards with the help of a boundary step:\\
When a Monge cone is intersected by a boundary, e.g. a shock wave, the
bicharacteristics are no longer drawn backwards in coordinate directions 
but in directions normal and tangential to the boundary (fig. 3). 
This improves the accuracy of the step considerably. 
The intersection points of the bicharacteristics with the boundary 
are determined, the footpoint values are interpolated at the boundary surface, 
and the grid point is updated.
The boundary step is only of first order because the lateral derivatives 
cannot be eliminated with the Butler procedure in case the bicharacteristics 
of a monge cone are cut off at different time levels.\\

\begin{figure}[h] 
\epsfclipon
\centerline{\epsfxsize 6cm \epsfbox{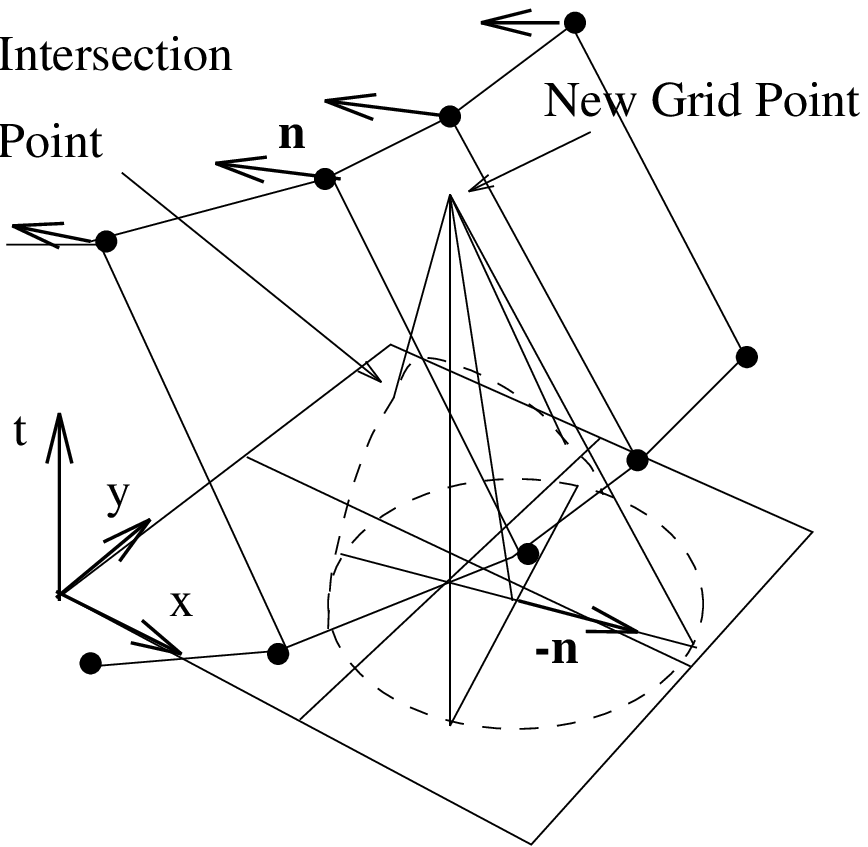}}
\centerline{Figure 3: Boundary Step}
\end{figure}

\vspace{1cm}

\noindent
{\bf \underline{Interpolation}}

In the vicinity of discontinuity surfaces the  interpolation algorithm used 
for ordinary grid points becomes ill-conditioned, producing instabilities. 
  The most preferable way of interpolation in this case proved to be 
  a two-dimensional, second order least-squares pattern, sized 7 $\times$ 7 
  grid points, in which all upstream grid points are ignored, but in which 
  all shock points of this area are taken into account. 
  According to our experience an interpolation in smaller regions will 
  render the scheme unstable. \\ 

\noindent
{\bf COMPARISON WITH OTHER NUMERICAL SCHEMES}\\

The bicharacteristic scheme was tested with an analytical solution, and
computations of converging shock waves were compared with the 
corresponding results of a UNO scheme and the Moretti scheme
(a $\lambda$-scheme working with shock fitting).
As expected, the bicharacteristic scheme turned out to be 
superior to the UNO scheme regarding the prediction of position and velocity 
of the shock front.
The shock fitting algorithm provides exact information on 
position, direction of propagation, and local Mach number of the shock wave.\\ 

\begin{figure}[h] 
\epsfclipon
\centerline{\epsfxsize 7cm \epsfbox{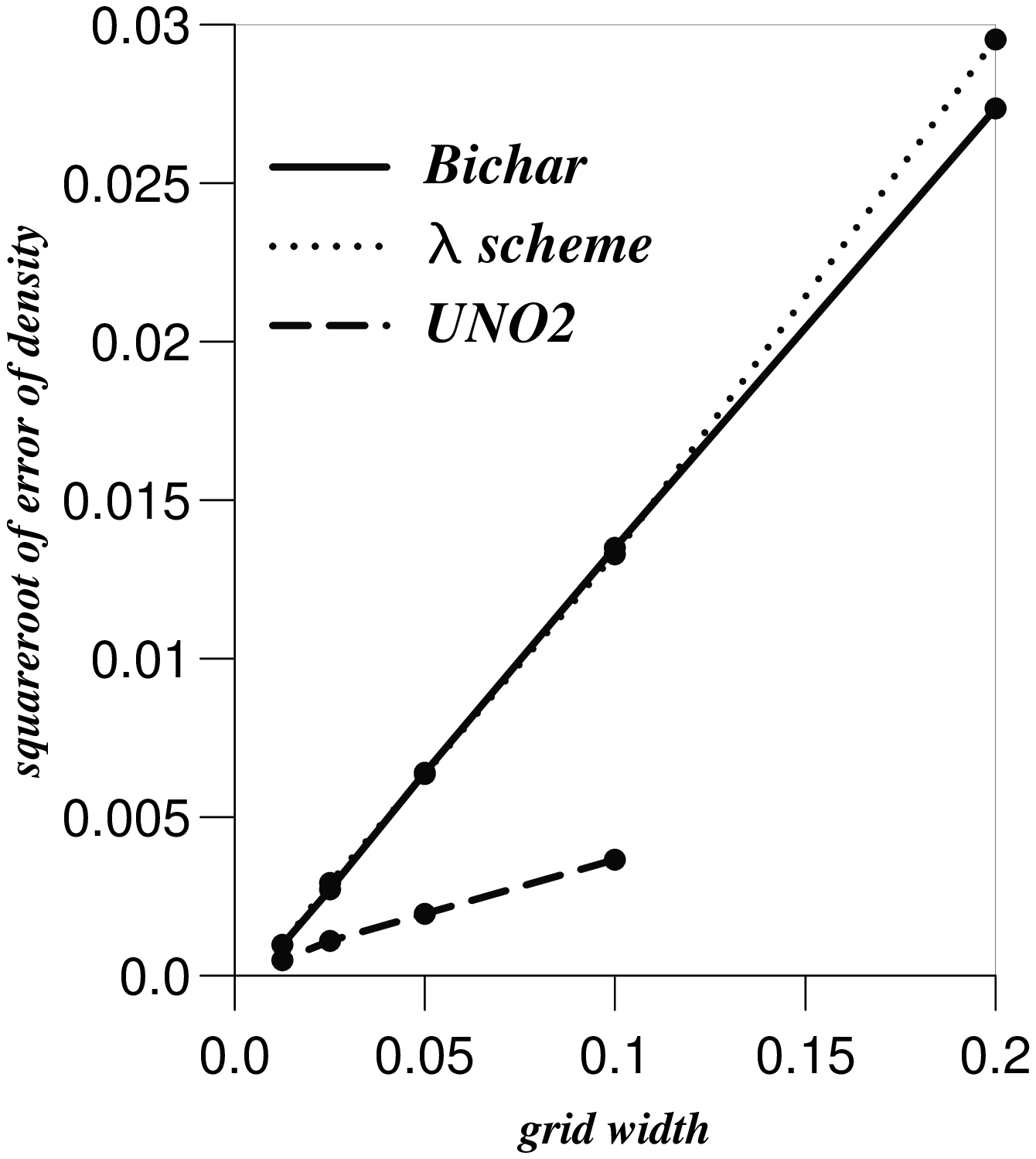}}
Figure 4: Comparision of convergence of the UNO-, $\lambda$- 
and bicharacteristic scheme
\end{figure}

The Moretti scheme [4] proved to be comparable in accuracy
to the bicharacteristic scheme (fig. 4). 
The aforementioned is faster (by a factor 1.5, assuming Courant 
Number 2.0). This is due to the 
time consuming interpolations in the bicharacteristic scheme.
We found that the convergence of the shock front position 
towards a known solution by refinement of the grid 
is slightly faster in the bicharacteristic scheme, due to its boundary step.
 
  Recently, Nasuti and Onofri [5] extended
  Moretti's original shock fitting algorithm to handle triple points.
  We implemented their extensions in our version 
  of the $\lambda$-scheme and added some further improvements 
  (to be published soon). 
  We found, that the shock fitting algorithm proposed by these authors
  still has some disadvantages. For example: 
  The fragmentation of the shock front,
  described in the next paragraph occurs too late and is not 
  enough pronounced
  in case the shock contour has an unfavourable orientation to the 
  computational grid;
  whereas our shock fitting algorithm is not influenced by the relative 
  position of shock front and computational grid.\\

\noindent
{\bf COMPUTATION OF CONVERGING SHOCK WAVES}\\
  
The computation shown here (fig. 5 and 6) started with a 
slightly deformed shock wave of Mach No. 2.5, at radius 1.0. 
Isopycnics of a time step immediately before and some time after 
fragmentation occurs are shown in Figure 5 and 6, respectively. 
Presently, the computation of converging shock waves extends 
to the instant of fully developed fragmentation, just before 
reflection of the leading shock. The extension of the 
method of bicharacteristics to proceed beyond this point 
is under work and does not pose any fundamental problems.\\

Calculating converging shock waves with a bicharacteristic scheme, one has
to make sure that the Mach stem will develop correctly. As mentioned above, 
information on the downstream values of the physical quantities reaches 
a shock point by transportation along a retrograde bicharacteristic.
The footpoint of this bicharacteristic generally is located only fractions 
of a grid width apart from the shock front.
However, the physical mechanism which generates the Mach stem is a density 
hump which gradually steepens as the shock wave converges. Finally, this 
density hump takes the form of a bow shock [6, 7]. 
Emerging gradually from a compression wave, its shock profile is smeared out
similar to shocks in common difference schemes over a number of grid points.
For this reason the retrograde bicharacteristic sees only a slight increasing
of e.g. density values where a marked density hump should be located.
Therefore, the local shock velocity could be calculated too small 
and, hence, the formation of the Mach stem could be delayed.

\begin{figure}[h]
\epsfclipon
\centerline{\epsfxsize 8cm \epsfbox{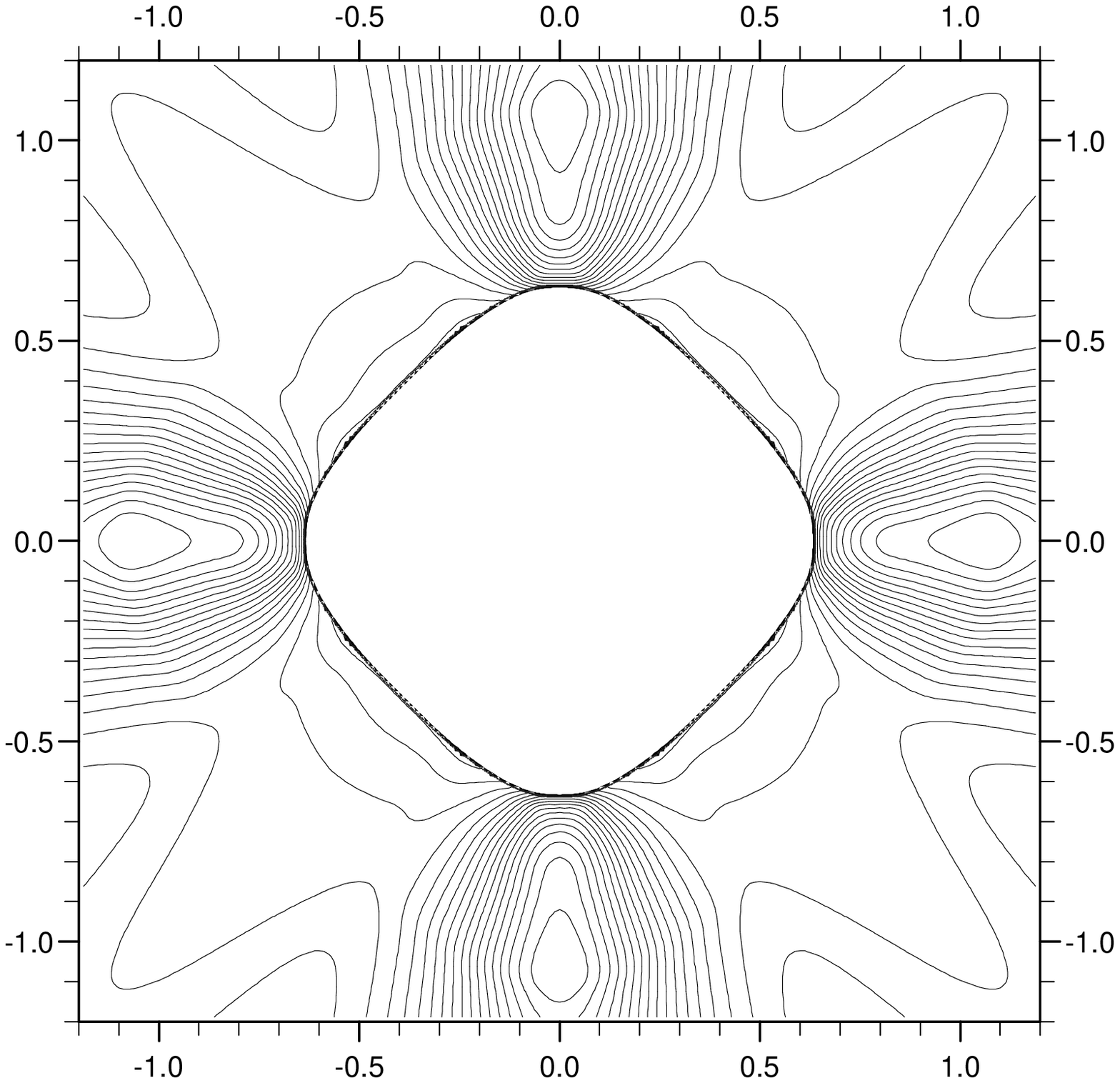}}
Figure 5: Isopycnic at t=0.1750 (before fragmentation)\\
The bold contour represents the leading shock wave.
\end{figure}

\begin{figure}[h] 
\epsfclipon
\centerline{\epsfxsize 8cm \epsfbox{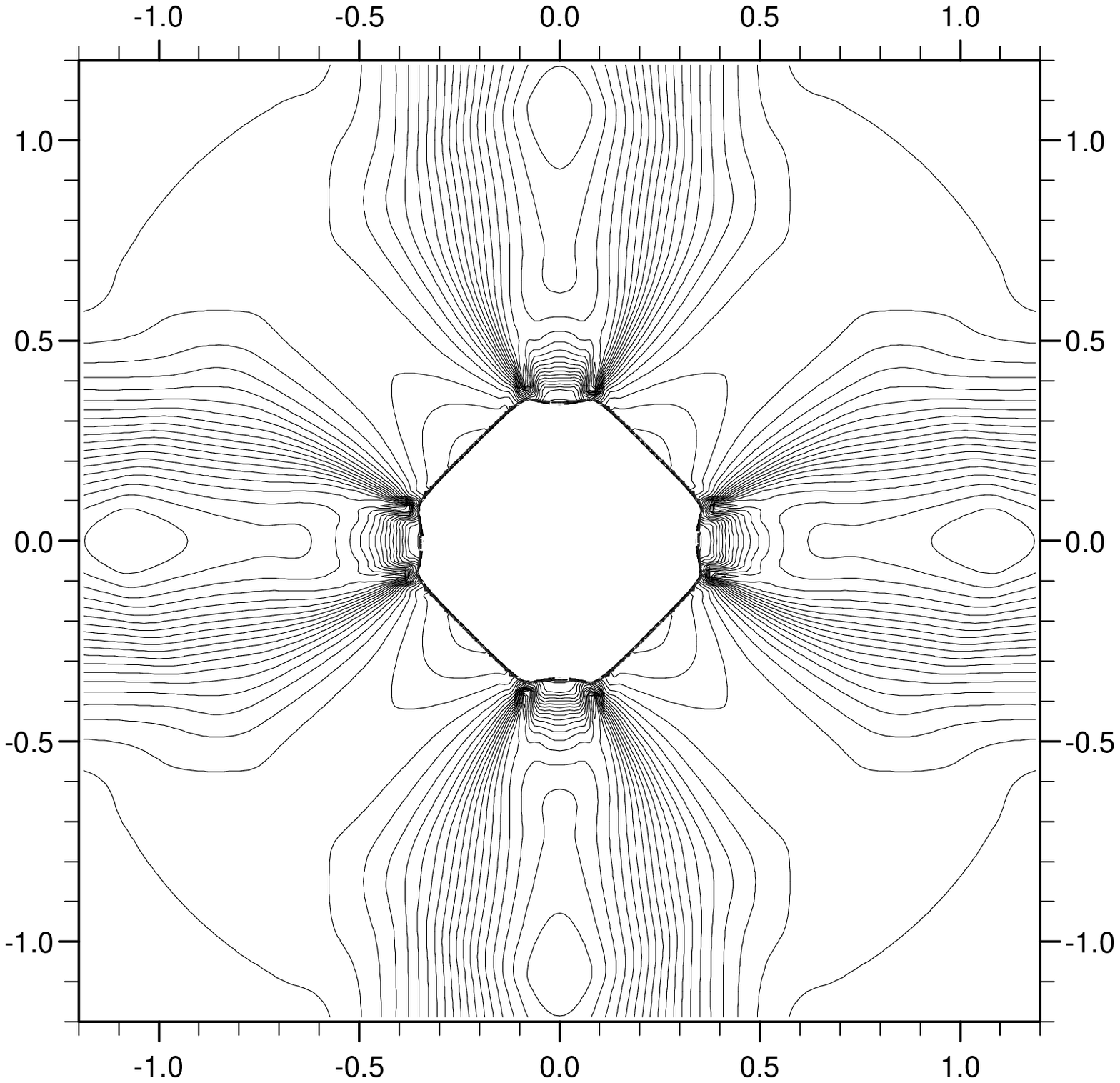}}
Figure 6: Isopycnic at t=0.2622 (after fragmentation).
The curvature of the Mach stem is below grid width 
and cannot be observed by shock capturing schemes in this resolution.  
Compare also fig. 7.
\end{figure}

To overcome this problem, the developing 
bow shock has to be detected with a pattern recognition algorithm, 
and consecutively be treated with the shock fitting algorithm.\\

As mentioned above, the density hump gives rise to 
new developing shocks in the vicinity of the triple points (fig. 7).
These lateral waves are not fitted yet because of the considerable 
effort, this might take. 
However, the computed solution is acceptable as long as the lateral waves 
remain sufficiently weak. 
Depending on our initialisation the Mach Number of the lateral waves 
was only slighly above 1.0 
and therefore the results can be considered as valid.\\

\vspace{2cm}

\noindent
{\bf OUTLOOK}\\

The following improvements seem feasible:\\
-The treatment of intersecting shock point trajectories should be 
reconsidered.\\
-Newly developing shock fronts and discontinuity lines should be 
detected and fitted as proposed by Moretti.\\

The bicharacteristic scheme with shock fitting presented here 
provides great accuracy and physical insight and allows a 
variety of applications:
It could be used to compute, e.g., channel flows,
flows around wings, MHD problems, star pulsation, and non-equilibrium flows.
The scheme can also profitably be used as a standard 
to estimate the capabilties of other schemes.\\

\begin{figure}[h]
\epsfclipon
\centerline{\epsfxsize 8cm \epsfbox{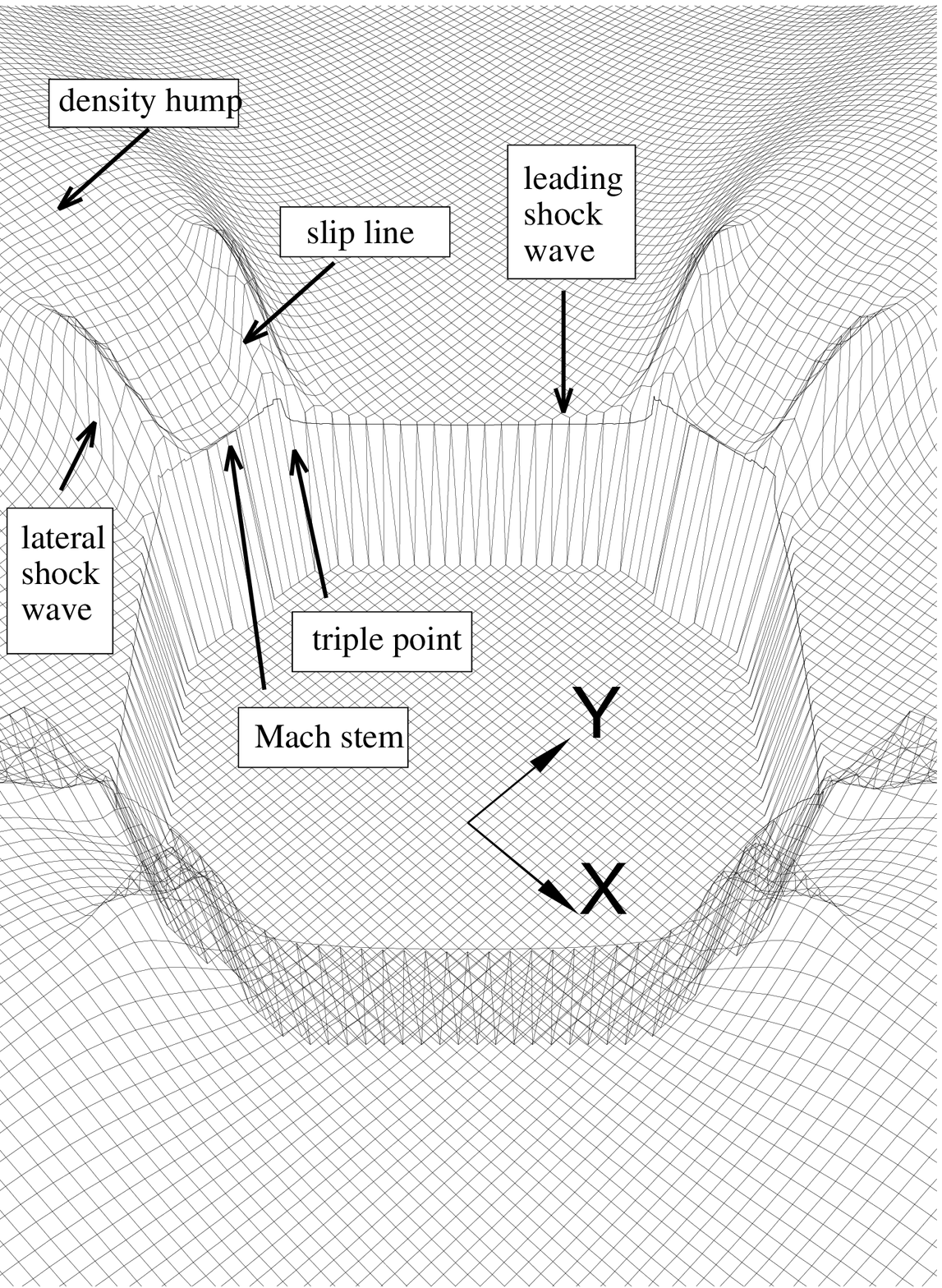}}
Figure 7: Density at t=0.2622\\
Due to the shock fitting algorithm the 
leading shock wave, which is represented here by density values 
exactly fullfilling the Rankine-Hugoniot relations,
is not smeared out.
The eminent area between lateral shock wave and slip line 
consists of those fluid particles
which passed both the leading shock wave and the lateral wave. 
\end{figure}

\noindent
{\bf NOMENCLATURE}\\

\vspace{-0.4cm}

\begin{tabbing}
\hspace{3.5cm}\=\kill
$a$                          \> sound speed\\
$\vec B$                     \> vector along a bicharacteristic\\
$\vec C_0$                   \> vector along the particle path\\
$D_t=\partial_t +\left(\vec v \nabla \right)$ \> substantial derivative\\
$D_0=\partial_t +\left(\vec v \nabla \right)$ \> derivative along $\vec C_0$\\
$D_B=(\vec B\nabla)$                              \> derivative along $\vec B$\\
$\partial_g=(\vec g\nabla)$ \> derivative in direction of $\vec g$\\
$\partial_t= \partial/\partial_t$   \>time derivative\\
$\vec g$                     \> unit vector (spatial)\\
$P=ln(p/p_0)$                \> logarithmic pressure\\
$p, p_0$                     \> pressure, initial upstream press.\\
$\vec v$                     \>velocity field\\
$v_g =\vec v\vec g$          \> projection of $\vec v$ on $\vec g$\\ 
$\gamma$                     \> specific heat ratio\\
$\varrho=ln(\rho/\rho_0)$    \> logarithmic density\\
$\rho$, $\rho_0$             \> density, initial upstream density\\
\end{tabbing}

\noindent
{\bf ACKNOWLEDGEMENTS}\\

We like to thank O. Rautenberg and A. v. Endt for implementation of the
Moretti Scheme, and H. Wallus for maintenance of the UNO2 program.\\

\noindent
{\bf REFERENCES\\}

\small
[1] Hartree, D. R., ''Some practical methods of using characteristics in the calculation of non-steady compressible flows''. US Atomic Energy Comm. Report AECU-2713 (1953)\\

[2] Butler, D. S., The numerical solution of hyperbolic systems of partial
differential equations in three independent variables, Proceedings of the Royal Society Vol. A255, (1960), p. 232-252\\ 

[3] Moretti, G., Detection and fitting of two-dimensional shocks, Notes Num. Fluid Mech. Vol 20, (1987), p. 239-246\\

[4] Moretti, G., Efficient Euler Solver with many applications,
AIAA Journal, Vol. 26 No. 6, (1988), p. 655-660\\ 

[5] Nasuti, F., Onofri, M., Analysis of Unsteady Supersonic Viscous Flows by a Shock-Fitting Technique, AIAA Journal, Vol. 34, No.7, (1996), p. 1428-1434\\

[6] Watanabe, M., Takayama, K., Stability of converging cylindrical shock waves, Shock waves, Springer Verlag, (1991),  p. 149-160\\

[7] Ben Dor, G. Shock Wave Reflection Phenomena, Springer Verlag, (1992)\\

\end{document}